\documentclass[12pt]{article}

\usepackage{amsfonts}

\def\Z{\mathbb{Z}}
\def\R{\mathbb{R}}

\def\Im{\mathrm{Im}}
\def\Re{\mathrm{Re}}
\def\vol{\mathrm{vol}}

\newcommand{\gsim}{ \mathop{}_{\textstyle \sim}^{\textstyle >} }
\newcommand{\lsim}{ \mathop{}_{\textstyle \sim}^{\textstyle <} }
\newcommand{\vev}[1]{ \left\langle {#1} \right\rangle }

\def\diag{\mathop{\rm diag}\nolimits}

\def\change#1#2{{#2}}

\begin{document}
\baselineskip 0.6cm

\begin{titlepage}

\begin{flushright}
UT-03-38 \\
RESCEU-51/03 \\
LBNL-54070 \\
\end{flushright}

\vskip 2cm
\begin{center}
{\large \bf Supergravity Analysis of \\ 
Hybrid Inflation Model from D3--D7 System }

\vskip 1.2cm
Fumikazu Koyama$^a$ Yuji Tachikawa$^a$ and Taizan Watari$^b$

\vskip 0.4cm
${}^a$
{\it Department of Physics, University of Tokyo, \\
          Tokyo 113-0033, Japan}\\

${}^b$
{\it Department of Physics, University of California at Berkeley, \\
          Berkeley, CA 94720, USA}\\

\vskip 1.5cm
\abstract{The slow-roll inflation is a beautiful paradigm, yet 
the inflaton potential can hardly be sufficiently flat 
when unknown gravitational effects are taken into account.
However, the hybrid inflation models constructed in D = 4 ${\cal N}$ = 1 
supergravity can be consistent with ${\cal N}$ = 2 supersymmetry, 
and can be naturally embedded into string theory. 
This article discusses the gravitational effects carefully 
in the string model, using D = 4 supergravity description.
We adopt the D3--D7 system of Type IIB string theory compactified on 
$K3 \times T^2/\Z_2$ orientifold for definiteness. 
It turns out that the slow-roll parameter can be sufficiently small 
despite the non-minimal K\"{a}hler potential of the model.
The conditions for this to happen are given in terms 
of string vacua.
We also find that the geometry obtained by blowing up singularity, 
which is necessary for the positive vacuum energy, 
is stabilized by introducing certain 3-form fluxes. }

\end{center}
\end{titlepage}


\section{Introduction}\label{sec:intro}

The slow-roll inflation is a beautiful paradigm, in which 
not only the flatness and homogeneity of the universe but also 
the origin of the scale-invariant density perturbation is understood.
However, it is not easy to obtain a scalar potential $V$ that 
satisfies the slow-roll conditions \cite{slow-roll}
\begin{equation}
 \eta \equiv \frac{M_{\rm pl}^2 V''}{V} \ll 1, \qquad 
 \epsilon \equiv \frac{1}{2}
         \left(\frac{M_{\rm pl}V'}{V}\right)^2 \ll 1,
\label{eq:sr}
\end{equation} 
where $V'$ and $V''$ are the first and second derivatives of $V$
with respect to the inflaton, and 
$M_{\rm pl}$ is the Planck scale $\simeq 2.4 \times 10^{18}$ GeV.
Suppose that there is vacuum energy $v_0^4$, and 
then one can see that even gravitational corrections to the potential 
\begin{equation}
 V(\sigma) = v_0^4 \left( 
   1 + c \left(\frac{\sigma}{M_{\rm pl}} \right)
     + c'\left(\frac{\sigma}{M_{\rm pl}} \right)^2 + \cdots 
                   \right)
\label{eq:grav-corr}
\end{equation}
are not allowed by the slow-roll conditions if the coefficients 
$c,c'$ are of the order of unity. 
Thus, the slow-roll inflation is sensitive even to the 
physics at the Planck scale, and can be a good probe 
in uncovering the fundamental laws of physics.

The hybrid inflation model \cite{Linde} is realized by quite 
simple models of D = 4 ${\cal N}$ = 1 supergravity (SUGRA) 
\cite{F-SUSY-a,F-SUSY-b,F-SUSY-c,D-SUSY}. 
Thus, the inflaton potential is protected from radiative corrections. 
However, D = 4 ${\cal N}$ = 1 SUGRA is not enough in \change{}{controlling}
the gravitational corrections.
In SUGRA as an effective-field-theory approach, 
no assumption except symmetries is imposed on ultraviolet physics.
Then, higher order terms are expected to be in K\"{a}hler potential 
with ${\cal O}(1)$ coefficients:
\begin{equation}
  K = X^\dagger X + k \frac{(X^\dagger X)^2}{M_{\rm pl}^2} + \cdots,
\end{equation}
where $X$ is a chiral multiplet containing the inflaton $\sigma$.
The second term contributes to the slow-roll parameter $\eta$, 
unless the vacuum energy is carried only by D-term.
Thus, the inflaton potential is not expected to be sufficiently 
flat. This is called the $\eta$ problem.

It is remarkable that the hybrid inflation model in ${\cal N}$ = 1 
supersymmetry (SUSY) is consistent with D = 4 ${\cal N}$ = 2 rigid 
SUSY \cite{WY}. The inflaton belongs to a vector multiplet of 
${\cal N}$ = 2 SUSY, and its interactions, including the K\"{a}hler 
potential, are highly constrained. 
Thus, it was argued in \cite{WY} that the ${\cal N}$ = 2 SUSY 
might ease the $\eta$ problem.
However, it was far from clear how the ${\cal N}$ = 2 SUSY can 
coexist with chiral quarks and leptons in D = 4 theories.

Superstring theory is a promising candidate of the quantum
theory of gravity. One can work out how the gravitational corrections 
look like, once a vacuum configuration is fixed. 
Thus, it is quite important 
in its own right to consider 
whether it can realize the slow-roll inflation.
Moreover, extended SUSY and higher dimensional spacetime 
are generic ingredients of string theory, and hence 
it is a plausible framework in \change{}{accommodating} the hybrid inflation
model with ${\cal N}$ = 2 SUSY; Enhanced ${\cal N}$ = 2 SUSY can 
coexist with other ${\cal N}$ = 1 supersymmetric sectors 
owing to the internal spacetime. 

It was shown in \cite{HHK} that the hybrid inflation model 
with ${\cal N}$ = 2 SUSY is realized by  D3--D7 system 
placed on a local geometry ALE $\times$ {\bf C}.
Thus, this framework of the Type IIB string theory enables us 
to examine if the inflaton potential can really be flat 
even when the internal dimensions are compactified and gravitational 
effects are taken into account.
Note that an analysis at the level of rigid SUSY, 
where $M_{\rm pl}$-suppressed corrections are neglected, 
is not sufficient to see the flatness of the inflaton potential.

This article is organized as follows. 
In section \ref{sec:setup}, we describe 
how the hybrid inflation model can be embedded in a local 
part of a realistic Calabi--Yau compactification 
of the Type IIB string theory. 
After that, we show that short-distance effects 
in the inflaton potential is not harmful, 
partly because of a translational invariance 
of the local geometry ALE $\times$ {\bf C},  
and partly because of a property specific to string theory.
In section \ref{sec:4DSUGRA}, we adopt $K3 \times T^2$ as 
a toy model of a Calabi--Yau 3-fold, 
and show in D = 4 SUGRA description that the potential 
is flat in the presence of dynamical gravity, consistent with 
the intuitive picture obtained in string theory. 
Special form of K\"{a}hler potential and interactions 
derived from string theory play crucial role there.
In section \ref{sec:moduli}, an explicit model that 
stabilizes non-zero Fayet--Iliopoulos parameters is given 
in subsection \ref{ssec:quaternionic}. 
The slow-roll parameter $\eta$ is evaluated for the model, 
and we obtain a condition that leads 
to the slow-roll inflation in subsection \ref{ssec:mixing}. 

We noticed that an article \cite{recent} was submitted to the e-print 
archive when we are completing this article. 
It has an overlap with this article in subjects discussed.

\change{}{Note added in version 2:}
There was an error (in identification of closed-string zero modes 
with fields in SUGRA) in the first version 
of this article, which was pointed out 
in \cite{recent2}\footnote{We are grateful 
to the authors of Ref.~\cite{recent2}.}.
It is corrected in this version, yet the main stream of 
logic (related to inflation) has not been changed 
from the first version.

\change{}{
Note added in version 3:
Wrong signs in eq.(42) were corrected, which requires a little 
modification in the model.
Corrections are limited in sections 4.1 and 4.2.
}
\section{String Theory Setup and \\ 
Short-Distance Effects in Inflaton Potential}\label{sec:setup}

The low-energy spectrum consists of an ${\cal N}$ = 2 SUSY 
vector multiplet $(X,V)$ when a space-filling fractional D3-brane 
is moving in ALE $\times$ {\bf C}.
The fractional D3-brane is regarded as a D5-brane wrapped on 
a 2-cycle of the ALE space \cite{DDG}, 
and hence is trapped at a tip of the ALE space. 
When a space-filling D7-brane is further introduced and stretched 
in the ALE direction, ${\cal N}$ = 2 SUSY is preserved, 
and one massless hypermultiplet $(Q,\bar{Q})$ arises 
from strings connecting the D3 and D7 branes.
D7--D7 open string and closed string are not dynamical degrees of 
freedom because of the infinite volume of ALE $\times$ {\bf C}.
The superpotential is given by 
\begin{equation}
 W = \sqrt{2}g (\bar{Q} X Q - \zeta^2 X), 
\label{eq:rigidN=2}
\end{equation}
and there may be Fayet--Iliopoulos D-term ${\cal L} = - \xi^2 D$.
The inflaton is $X$, which corresponds to the distance 
between the D3 and D7-branes in the {\bf C}-direction.
When the D3-brane becomes close enough to the D7-brane, i.e., 
$X \lsim |\zeta|,\xi$, the D3--D7 open string modes $(Q,\bar{Q})$ 
become tachyonic and begin to condense, 
a D3--D7 bound state is formed, the vacuum energy 
$g^2/2 \times (|2 \zeta^2|^2+\xi^4)$ disappears, 
and the inflation comes to an end. 
There is no massless moduli in this vacuum, 
and this is the reason why the fractional brane is adopted.
The Fayet--Iliopoulos parameters 
$(-2 \Im \zeta^2,2 \Re \zeta^2, \xi^2)$ are non-zero 
when a singularity ${\bf C}^2/\Z_M$ is blown up 
to be a smooth ALE space \cite{DM}\footnote{
See also \cite{DHHK}, where the vacuum energy is given by 
the vacuum expectation value of the B field.}.

Type IIB string theory has to be compactified on a Calabi--Yau 
3-fold in order to obtain dynamical gravity.
The D7-brane should be wrapped on a homomorphic 4-cycle so that 
the D = 4 ${\cal N}$ = 1 SUSY is preserved \cite{BBS}.
We consider that there is a point on the 4-cycle around which 
the local geometry of the Calabi--Yau 3-fold is ALE $\times$ {\bf C}.
The fractional D3-brane is trapped at the tip of the ALE space, 
and is able to move along the ${\bf C}$-direction. On the other hand, 
${\cal N}$ = 1 vector multiplet is usually the only Kaluza--Klein 
zero mode from the D7--D7 open string, and in particular, 
the coordinate of the D7-brane in the ${\bf C}$-direction is fixed.
Other particles such as quarks and leptons can be realized 
by local construction of D-branes at another place 
in the Calabi--Yau 3-fold, as in \cite{Madrid}.
Thus, the non-compact model above can be embedded as a local model 
of a realistic Calabi--Yau compactification.

The world-sheet amplitude of string theory is expanded 
in powers of the string coupling $g_s$. 
The expansion begins with the sphere amplitude, 
which is proportional to $g_s^{-2}$. 
In particular, $M_{\rm pl}^2$ is proportional 
to $g_s^{-2}$.

The disc amplitude comes at the next-to-leading order, $g_s^{-1}$. 
It is calculated by restricting the boundary of the world sheet 
to the fractional D3-brane. 
The kinetic term of the inflaton arises at this level, and hence 
its coefficient is proportional to $g_s^{-1}$. 
The kinetic term of the U(1) vector field, the ${\cal N}$ = 2 SUSY 
partner of the inflaton, also has a coefficient proportional to 
$g_s^{-1}$.
Thus, the U(1) gauge coupling constant $g$ is related to $g_s$ via 
$g^2 \sim g_s$.
The vacuum energy also arises at this level. 
Therefore, the vacuum energy is proportional to $g_s^1 \sim g^2$
when $M_{\rm pl}^2 \sim g_s^{-2}$ is factored out 
from the scalar potential 
(see also the discussion at the end of this section). 

We are interested only in the disc amplitude whose boundary is on the 
D3-brane. 
The D7-brane is irrelevant, and only the local background geometry 
around the D3-brane, ALE $\times$ {\bf C}, is relevant to the 
disc amplitude. 
Since ALE $\times$ {\bf C} has translational invariance 
in the {\bf C}-direction, 
the translational invariance is respected in the disc amplitude.
Thus, the amplitude does not depend on the position of the D3-brane.
Therefore, the disc amplitude does not induce inflaton potential.

The cylinder amplitude is at the next order, $g_s^0$.
The 1-loop amplitude of open string and the amplitude 
exchanging closed string at the tree level are contained here.   
The inflaton potential comes from a cylinder with one end 
on the D3-brane and the other on the D7-branes.
The amplitude contains a potential logarithmic in the distance $r$ 
between the two D-branes. This potential corresponds 
to the 1-loop radiative correction in \cite{F-SUSY-b}. 
There are also terms damping exponentially in $r$.
They are interpreted as the forces between the two D-branes 
induced by exchanging stringy excited states at the tree level. 
These terms are suppressed very much when 
the D-branes are separated by a distance longer than 
the string length $\sim \sqrt{\alpha'}$.
Finally, there is also a term quadratic in the inflaton $r$. 
This potential is induced by exchanging massless twisted 
sector fields; both the fractional D3-branes and the D7-brane  
carry twisted Ramond--Ramond charges.

Putting all above together, we have obtained 
\begin{equation}
 {\cal L} \sim  \left(g_s^{-2} \sim M_{\rm pl}^2 \right) 
                \left( R + g_s (\partial r)^2 
    - g_s \left(1 + g_s \ln r - g_s e^{-r} + g_s r^2 + \cdots 
          \right)
                          \right),
\end{equation}
where $\alpha'$ is set to unity and $r$ is the distance between 
the two D-branes. 
Let us now rescale the inflaton $r$ so that 
the kinetic term is canonical; 
$\sigma \equiv \sqrt{g_s} r M_{\rm pl}$. 
Then, the scalar potential is given by 
\begin{equation}
 V \propto g_s M_{\rm pl}^2 
      \left( 1 + g_s \ln \left(\frac{\sigma}{M_{\rm pl}}\right) 
               - g_s e^{
                      -\frac{\sigma}{\sqrt{g_s}M_{\rm pl}} 
                        }
               + \left(\frac{\sigma}{M_{\rm pl}}\right)^2 
               + {\cal O}(g_s) 
                    \left(\frac{\sigma}{M_{\rm pl}}\right)^2 
               + {\cal O}\left( 
                    \left(\frac{\sigma}{M_{\rm pl}}\right)^3 
                         \right)
      \right).
\label{eq:crude}
\end{equation}
The correct mass dimension of the scalar potential is restored 
by multiplying quantities that have been set to unity, 
including $\alpha'$ and the volume of the compactified manifold.
Note that the short-distance effects appear only 
as the exponentially damping potential. 
This is partly because the local translational invariance 
of the internal space dimensions forbids the potential 
from the disc amplitude.
This is also because the cylinder amplitude is
interpreted as the Yukawa potential induced by heavy states, 
and hence the short-distance (ultraviolet) effects is irrelevant 
unless the D3-brane is in a short distance from the D7-brane 
in the internal space dimensions.
This kind of picture is hardly obtained without 
assuming string theory.
The logarithmic correction is not harmful when the coupling is
sufficiently small, just as in field theoretical models 
\cite{F-SUSY-b}.
The quadratic potential induced by the twisted-sector exchange, 
which can be the only harmful effect, is suppressed in certain 
string vacua as shown in section \ref{sec:moduli}.
Although the volume of the Calabi--Yau 3-fold has not been treated 
carefully, it is also shown in section \ref{sec:4DSUGRA} 
and \ref{sec:moduli} that this parameter is irrelevant 
to the flatness of the inflaton potential.


\section{D = 4 SUGRA Analysis 
of the Inflaton Potential}\label{sec:4DSUGRA}


Both the Planck scale and the Kaluza--Klein scale are finite, 
as well as the string scale, when the internal dimensions 
are compactified.
We show in this section that the inflaton potential 
still reflects the translational invariance 
of the local geometry, and is sufficiently flat, 
even in the low-energy effective 
D = 4 SUGRA description obtained after the compactification.
In particular, the inflaton potential does not grow exponentially 
for large field value, even when the vacuum energy is carried by 
F-term.
It is another purpose of this section and of section
\ref{sec:moduli} to examine the volume-parameter (in)dependence 
of the potential, which was neglected in the previous section.  

We adopt $K3 \times T^2$ as the model of a Calabi--Yau 3-fold.
It surely contains ALE $\times$ {\bf C} as a local geometry, but
it also preserves extended SUSY.
Thus, the analysis based on $K3 \times T^2$ 
has a limited meaning.
However, this toy model has another virtue that we can analyze 
more precisely owing to the extended SUSY.
Furthermore, a related discussion is found at the end of this section.

The scalar potential of the D = 4 ${\cal N}$ = 2 SUGRA is given by 
\cite{N2SUGRA}
\begin{equation}
 V = 4 h_{uv}k^u_\Lambda k^v_\Sigma L^\Lambda L^{* \Sigma}
   + \left(g^{ij^*} f_i^\Lambda f_{j^*}^\Sigma 
          - 3 L^{*\Lambda} L^\Sigma\right)P^x_\Lambda P^x_\Sigma.
\label{eq:fullpotential}
\end{equation}
$P^x_\Lambda$ are momentum maps, 
which roughly correspond to D-term (Killing potential) 
and F-term potential,  
$k^u_\Lambda$ Killing vectors, 
$L^\Lambda$ is roughly the scalar partner of the $\Lambda$-th vector field, 
$f^\Lambda_i$ its covariant derivative with respect 
to the $i$-th scalar of the vector multiplets, 
$g_{ij^*}$ and $h_{uv}$ the metric of vector- and hypermultiplets, 
respectively.
See \cite{N2SUGRA} for more details.  

Let us define 
\begin{eqnarray}
 L^\Lambda & \equiv & e^{\frac{K_V}{2}} X^\Lambda, \label{eq:LX}\\
 W_0 & \equiv & X^\Lambda (P^1 + iP^2)_\Lambda,
\end{eqnarray}
where $K_V$ is the K\"{a}hler potential of vector multiplets.
Then, the first term of (\ref{eq:fullpotential}) 
becomes $e^{K_V} |\partial W_0|^2 $ 
for hypermultiplets, and the second contains 
$e^{K_V}|\partial W_0|^2$ for ${\cal N}$ = 1 chiral components 
of ${\cal N}$ = 2 vector multiplets.
The last term contains $-3 e^{K_V} |W_0|^2$.
Thus, the ${\cal N}$ = 2 SUGRA scalar potential is not completely
different from that of ${\cal N}$ = 1 SUGRA.  
See \cite{2to1} for more details about the relation between 
${\cal N}$ = 2 SUGRA and ${\cal N}$ = 1 SUGRA. 
We revisit this issue at the end of this section.

${\cal N}$ = 1 chiral multiplet $X$ in section \ref{sec:setup}, 
identified with the inflaton, belong to an ${\cal N}$ = 2 vector 
multiplet.
Thus, one of $X^\Lambda$'s is approximately $X$. 
The ${\cal N}$ = 2 hypermultiplet $(Q,\bar{Q})$ 
in section \ref{sec:setup}
are in the momentum maps as 
\begin{eqnarray}
 P^3_\Lambda & = & \vev{e^3} + |Q|^2 - |\bar{Q}|^2 + \cdots ,
     \label{eq:D3Dterm}\\
i (P^1+iP^2)_\Lambda & = &  i \vev{e^1+ie^2} + 2 Q\bar{Q} + \cdots. 
     \label{eq:D3Fterm}
\end{eqnarray}
The Fayet--Iliopoulos parameters are now obtained as 
vacuum expectation values (VEV's) $\vev{e^m}$'s ($m=1,2,3$) 
of massless fields in the closed string sector; 
$i\vev{e^1+ie^2} = - 2 \zeta^2$ and $\vev{e^3} = \xi^2$.
The first term of (\ref{eq:fullpotential}) contains 
\begin{equation}
 g^2 (|X Q|^2 + |X\bar{Q}|^2),
\label{eq:hypermass}
\end{equation}
which prevents the D3--D7 open string modes $(Q,\bar{Q})$ 
from condensing during the inflation because $\vev{X}$ is large.
The vacuum energy (and the inflaton potential) during the inflation 
is (are) provided by the last two terms
\begin{equation}
 \left(g^{ij^*} f_i^\Lambda f_{j^*}^\Sigma 
          - 3 L^{*\Lambda} L^\Sigma\right)\vev{P^x_\Lambda P^x_\Sigma}, 
\label{eq:(U-3LL)PP}
\end{equation}
as we see explicitly in this section.
Although the first term also contributes to the inflaton potential, 
we show in section \ref{sec:moduli} that 
this contribution is negligible in certain string vacua.

Let us suppose that the inflaton potential comes dominantly from 
(\ref{eq:(U-3LL)PP}). 
Then, we only have to know the special geometry, which determines 
$g^{ij^*} f_i^\Lambda f_{j^*}^\Sigma  - 3 L^{*\Lambda} L^\Sigma$, 
to see whether the inflaton potential is flat. 
Therefore, we just assume  in this section that the positive 
$\vev{P^x_\Lambda P^x_\Sigma}$ is realized, and postpone 
discussing how the momentum maps are determined until 
section \ref{sec:moduli}.
Subsection \ref{ssec:quaternionic} discusses 
how to stabilize non-zero 
$\vev{e^m}$'s in (\ref{eq:D3Dterm}) and (\ref{eq:D3Fterm}) 
by examining the quaternionic geometry of hypermultiplets. 
Subsection \ref{ssec:mixing} explains when the first term 
in (\ref{eq:fullpotential}), which contains the quadratic term 
in (\ref{eq:crude}), is not harmful to the slow-roll condition.

\subsection{Special Geometry 
of the Vector Multiplets \\ and Calabi--Visentini Basis}\label{ssec:special}

We begin by determining the K\"{a}hler metric of 
the moduli space of vector multiplets (special geometry). 
After that, a symplectic vector $(X^\Lambda,F_\Sigma)$ is 
chosen suitably and  
$(g^{ij^*} f_i^\Lambda f_{j^*}^\Sigma  - 3 L^{*\Lambda} L^\Sigma)$  
in the potential (\ref{eq:(U-3LL)PP}) is calculated.

As we see later, one cannot capture the essential reason of 
the flatness in this SUGRA analysis without considering carefully 
the interaction of the inflaton with other vector multiplets 
arising from the closed string sector.
There are three ${\cal N}$ = 2 vector multiplets in the low-energy 
effective theory when Type IIB theory is compactified on 
$K3 \times T^2/\Z_2$. Here, $\Z_2$ is generated by 
$\Omega (-1)^{F_L} R_{T^2}$, where $R_{T^2}$ reflects 
the coordinates of $T^2$.
The three complex scalars in these multiplets are denoted 
by $S$, $T$ and $U$; $S=C_{(0)}+ig_s^{-1}$,
$\Im T \propto g_s^{-1}{\rm vol}(K3)$, and 
$U$ is the complex structure of $T^2$.
We adopt a convention in which imaginary parts of all 
$S$, $T$ and $U$ are positive.

The kinetic terms of these fields are determined from \cite{ABFPT}, 
since a model T-dual to ours 
(Type I theory compactified on $K3 \times T^2$) is discussed there.
We take the T-duality transformation from \cite{ABFPT}, and find that 
the kinetic term is given by 
\begin{equation}
\frac{\partial_\mu  S\partial_\mu \bar S}{(S - \bar S)^2}
+\frac{\partial_\mu  T\partial_\mu \bar T}{(T - \bar T)^2}
+\frac{\partial_\mu  U\partial_\mu \bar U}{(U - \bar U)^2}
\end{equation}
after Kaluza--Klein reduction and Weyl rescaling.
All the scalar fields are chosen to be dimensionless, 
and these terms become a part of D = 4 Lagrangian 
when multiplied by $M_{\rm pl}^2$.  
This metric of the special geometry, which is the target space 
of the non-linear $\sigma$ model of the scalar components,  
is obtained from a K\"{a}hler potential 
\begin{equation}
K_V=-\log \left( i (S - \bar S)(T - \bar T)(U - \bar U) \right),
\end{equation}
which can be derived from a prepotential 
\begin{equation}
{\cal F}=- STU.
\end{equation}

Let us now introduce D3-branes to this system. 
The coordinates of the D3-branes on $T^2$ are denoted by  
$(x_i,y_i) \sim (x_i+1,y_i) \sim (x_i,y_i+1)$. 
We introduce a complex scalar $Z_i = x_i + U y_i$.
The twisted Ramond--Ramond (RR) charge does not vanish 
when there is only one fractional D3-brane.
But, the RR-charge can be cancelled in a system where
D7-branes and other fractional D3-branes are introduced.
They will be scattered at different points in $T^2$. 
We are interested in only one\footnote{
Since we are interested only in the disc-level potential in this
section, other D-branes are irrelevant to the inflaton potential.} 
of the fractional D3-branes $Z=Z_1$, 
which corresponds to $X$ in section \ref{sec:setup}.

The kinetic terms of the bulk particles and the D3-brane 
are given by \cite{ABFPT}
\begin{equation}
\frac{\partial_\mu  S\partial_\mu \bar S}{(S - \bar S)^2}
+\frac{|\partial_\mu  T +(x\partial_\mu  y - y\partial_\mu  x)/2 |^2}{(T- \bar T)^2}
+\frac{\partial_\mu  U\partial_\mu \bar U}{(U - \bar U)^2}
+\frac{(\partial_\mu x + U \partial_\mu y) 
       (\partial_\mu x + \bar{U} \partial_\mu y) }
      {(U - \bar U)(T - \bar T)}
\label{KINBOSON}
\end{equation}
after Kaluza--Klein reduction\footnote{The relative normalization 
between the bulk particles ($S$, $T$, $U$) and the D-brane $Z$ 
is not precise. It turns out, however, that the slow-roll parameter  
$\eta$ is independent of the normalization. Thus, we do not pay
attention to the numerical coefficients, say, of the last term, 
very much in this article.}. 
The cross term in the kinetic term of $T$ has its origin in 
the Wess--Zumino term on the D-branes 
\begin{equation}
 \int_{D3} C^{(4)}_{\mu \nu x y} (\partial_\rho x) (\partial_\sigma y)
     dx^\mu \wedge dx^\nu \wedge dx^\rho \wedge dx^\sigma 
  = - \frac{1}{2} 
       \int d^4x (\partial_\rho C^{(4)}_{\mu\nu xy})
       \epsilon^{\mu\nu\rho\sigma} 
       (x \partial_\sigma y - y \partial_\sigma x).
\end{equation}

Now, a new coordinate 
\begin{equation}
 	\tilde T= T + \frac12 y_i Z_i
\end{equation}
is introduced, and $\tilde T$ is regarded as one of 
the special coordinates; $T$ is no longer a special coordinate. 
The K\"{a}hler potential for the metric (\ref{KINBOSON}) is given by 
\begin{eqnarray}
 K_V & = & -\log\left( i (S - \bar S)
   ((\tilde T - \bar {\tilde T})(U-\bar U) - (Z - \bar Z)^2/2 )\right) 
  \label{eq:Kahler}\\
   & = & -\log\left( i (S-\bar S)( T - \bar T)(U - \bar U) \right), 
  \label{eq:Kahler2} 
\end{eqnarray}
and this K\"{a}hler potential is derived from a prepotential 
\begin{equation}
{\cal F} =  - S\tilde TU + SZ^2/2. 
\end{equation}
Thus, newly introduced $\tilde T$ is in the correct set of 
special coordinates, along with $S,U$ and $Z$.
Note that the complexified coupling of the gauge field 
on the D3-brane is $S$, as desired.
The special geometry obtained here turns out to be 
\begin{equation}
\frac{SU(1,1)}{SO(2)} \times \frac{SO(2,3)}{SO(2)\times SO(3)}.
\end{equation}
One of the special coordinates $S$, which factorizes 
in (\ref{eq:Kahler}), parametrizes SU(1,1)/SO(2).

The symplectic section $\Omega = (X^\Lambda, F_\Lambda)$ 
of the special manifold is given by
\begin{eqnarray}
X^\Lambda &=& (1, S, \tilde T, U ,Z), \\
F_\Lambda &=& (S\tilde TU-SZ^2/2 , -\tilde TU +Z^2/2, -SU, 
               - S\tilde T, SZ).
\end{eqnarray}
Although the symplectic transformation of $\Omega$ does not change 
the K\"{a}hler potential, different choice of basis leads 
to different coupling with hypermultiplets \cite{AdAFP}. 
We choose a base in which 
the bi-doublet representation 
of SU(1,1) acting on $S$ and 
SU(1,1) $\subset$ SO(2,2) $\subset$ SO(2,3) acting on $U$ 
is realized in the coordinates $X^\Lambda$.
This is for the same reason as in \cite{AdAFL,recent2}.
Choosing a suitable symplectic transformation, 
one finds that 
\begin{eqnarray}
X^\Lambda&=& \left( \frac{1-SU}{\sqrt2},
	 -\frac{S+U}{\sqrt2}, 
         \frac{-1-SU}{\sqrt2},
         \frac{-S+U}{\sqrt2} ,Z \right), \label{eq:CV1}\\ 
F_\Lambda&=& \left( - \frac{\tilde T(1-SU)+SZ^2/2}{\sqrt{2}},
               - \frac{\tilde T(-S-U)+Z^2/2}{\sqrt{2}}, \right.  
                              \nonumber \\
         & & \left. \qquad \qquad 
               - \frac{\tilde T(1+SU)-SZ^2/2}{\sqrt{2}},
               - \frac{\tilde T(S-U)+Z^2/2}{\sqrt{2}},
                SZ \right).\label{eq:CV2}
\end{eqnarray}
This is the so-called Calabi-Visentini basis.

Now that we have holomorphic symplectic section 
$\Omega=(X^\Lambda,F_\Sigma)$ in a suitable basis, 
it is straightforward to calculate the potential (\ref{eq:(U-3LL)PP}). 
One finds that 
\begin{eqnarray}
\left(g^{ij^*} f_i^\Lambda f_{j^*}^\Sigma
- 3 e^{K_V} X^{*\Lambda} X^\Sigma \right) P^x_\Lambda P^x_\Sigma & = & 
- \eta^{\Lambda\Sigma}\frac{1}{2 \Im T} P^x_\Lambda P^x_\Sigma
|_{\Lambda,\Sigma=0,\ldots,3} 
 + \frac{1}{2 \Im S}P^x_{\Lambda=4}P^x_{\Lambda=4} \nonumber \\
 & & \!\!\!\!\!\!\!\!\!\!\!\!\!\!\!\!\!\!\!\!\!\!\!\!\!\!\!\!\!\!\!
     \!\!\!\!\!\!\!\!\!\!\!\!\!\!\!
     - \frac{x}{\sqrt{2} \Im T}
            (P^x_{\Lambda=0}+P^x_{\Lambda=2})P^x_{\Lambda=4}
     + \frac{y}{\sqrt{2} \Im T}
            (P^x_{\Lambda=1}+P^x_{\Lambda=3})P^x_{\Lambda=4}  ,
\label{eq:result}
\end{eqnarray}
where $\eta^{\Lambda \Sigma} = \diag(1,1,-1,-1)$.
The K\"{a}hler potential (\ref{eq:Kahler}) is far from minimal, 
and holomorphic symplectic section $\Omega$ 
in (\ref{eq:CV1}, \ref{eq:CV2}) exhibits 
intricate mixture of the special coordinates. 
However, the inflaton potential (\ref{eq:result}) 
is completely independent of the inflaton field $Z$, 
when $\vev{P_4 P_4}$ is non-zero.
This result shows that the flat inflaton potential is not lifted 
when the internal dimensions are compactified and 
the Planck scale (as well as the string scale) becomes finite.
See also section \ref{sec:moduli} for discussion related to 
the second line, which depends linearly on the inflaton.

The translational symmetry in the {\bf C}-direction, or in
$T^2$-direction, is preserved in the kinetic term of the bosons 
(\ref{KINBOSON}), 
where a scalar $\Re T$ from Ramond--Ramond 4-form potential 
are also shifted:
\begin{eqnarray}
	x\to x+\epsilon,& \qquad & \Re T\to \Re T -\epsilon y/2, \\
	y\to y+\epsilon',& \qquad & \Re T\to \Re T +\epsilon' x/2, 
\end{eqnarray}
or in terms of the special coordinates 
\begin{eqnarray}
	Z\to Z+\epsilon,& \qquad & 
        (\tilde TU-Z^2/2)\to (\tilde TU-Z^2/2) -\epsilon Z, \\
	Z\to Z+\epsilon' U, & \qquad & 
        \tilde T \to \tilde T+\epsilon' Z,
\end{eqnarray}
The translational symmetry of $T^2$ is now part of 
$SO(2,3)$ isometry 
along with $SO(2,2) \simeq SL_2 \R \times SL_2 \R$.

There is another interesting feature in (\ref{eq:result}).
Notice that F-term and D-term scalar potential are completely 
different in D = 4 ${\cal N}$ = 1 SUGRA, namely, 
\begin{equation}
 V_F = e^{K_V + K_H} \left(g^{ij^*}{\cal D}_i W_1 {\cal D}_{j^*} W_1^* 
                           - 3 |W_1|^2
                           \right), 
  \qquad V_D = g^2 |D|^2.
\label{eq:N1potential}
\end{equation}
However, the ${\cal N}$ = 2 scalar potential (\ref{eq:result}) 
``becomes''\footnote{Here, we keep quotation marks because 
there is a subtlety in defining K\"{a}hler potential $K_H$
for quaternionic geometry. See \cite{Louis} for more details.} 
\begin{equation}
 V_{x=1,2} = e^{K_V}\left(g^{ij^*}{\cal D}_iW_0 {\cal D}_{j^*}W_0^* 
                     - 3 |W_0|^2 \right) 
   ``=\mbox{''} V_F {\rm ~in~}(\ref{eq:N1potential}),
\end{equation}
where 
\begin{equation}
 W_1 ``=\mbox{''}
  e^{-\frac{K_H}{2}}W_0,
\end{equation}
while 
\begin{equation}
 V_{x=3} = g_s |P^3|^2 = V_D {\rm ~in~}(\ref{eq:N1potential}), 
\end{equation}
when the relation (\ref{eq:result}) holds.
Thus, the flat potential obtained in (\ref{eq:result}) 
may still be expected when the internal manifold is not 
$K3 \times T^2$ but a Calabi--Yau 3-fold 
with local ALE $\times$ {\bf C} geometry. 
Then, an important consequence is that the inflaton potential 
is not growing up exponentially at large field value, no matter 
how much the vacuum energy is carried by F-term in realistic models.

\section{Moduli Stabilization and 
Slow-roll Conditions}\label{sec:moduli}

In the previous section, we assumed that 
$\vev{P^x_{\Lambda=4} P^x_{\Sigma=4}}$ is non-zero. 
It is, however, realized as VEV's of dynamical fields, and  
would have vanished if those fields were not stabilized. 
Thus, we need to ensure that the non-zero VEV's of the dynamical 
fields are stabilized.

It has been clarified \cite{PS,Michelson} 
that most of moduli are stabilized by introducing 3-form fluxes.  
Moduli that are not stabilized by the 3-form fluxes can also be 
stabilized by non-perturbative effects.
Thus, it is not the main focus of our attention 
whether moduli are stabilized or not. 
Rather, the question is whether the stabilized Fayet--Iliopoulos 
parameter can be non-zero. 

Another important aspect of the moduli stabilization 
in models of inflation is that an extra inflaton potential 
is generically generated when stabilized heavy moduli 
are integrated out. 
Since even Planck-suppressed corrections are harmful 
to the flatness of the inflaton potential, extra contributions to the 
potential are also harmful when they are suppressed 
by masses of moduli. 
It also happens that the stabilizing potential 
sometimes constrains moduli as functions of the inflaton.
Thus, VEV's of moduli can change during the inflation, and 
the dynamics of the inflation can be different from 
the ordinary one. 
Therefore, the moduli stabilization is an important ingredient 
of the inflation model in string theory \cite{Toward}.

One can analyze the effects of introducing the fluxes 
in terms of D = 4 gauged SUGRA 
 \cite{Michelson}. 
We adopt $K3 \times T^2$ as a toy model of the Calabi--Yau 3-fold 
in this section (except in subsection \ref{ssec:CY}), 
to see explicitly how the non-zero 
Fayet--Iliopoulos parameters are stabilized and how the inflaton 
is mixed with other moduli.

The kinetic term of the Ramond--Ramond 4-form potential and the
Chern--Simons term are 
\begin{equation}
 \int d^{10}x 
  \frac{1}{2}
  \left|d C^{(4)} - \frac{1}{2} C^{(2)} \wedge dB 
                  + \frac{1}{2} B \wedge dC^{(2)} \right|^2 
+ \int C^{(4)} \wedge d B \wedge d C^{(2)}
\end{equation}
in D = 10 action of the Type IIB theory.
When the Type IIB theory is compactified on $K3 \times T^2/\Z_2$, 
the dimensional reduction of this action contains 
\begin{equation}
 \int d^4 x   \frac{1}{2}
       \left|\partial_\mu \int_{\Sigma \gamma \gamma'} C^{(4)} 
              + \frac12 \int_{\Sigma \gamma'} \vev{d B} 
                \int_{\gamma}C^{(2)}_{\gamma \mu} 
              - \frac12 \int_{\Sigma \gamma'} \vev{d C^{(2)}} 
                \int_{\gamma}B_{\gamma \mu} 
       \right|^2,
\label{eq:Stuckelberg}
\end{equation}
where $\Sigma$ denote 2-cycles of the $K3$ manifold and 
$\gamma,\gamma'$ 1-cycles of $T^2$. 
The quantities $\int_{\Sigma \gamma'} \vev{d B}$ and 
$\int_{\Sigma \gamma'}\vev{dC^{(2)}}$ are the number of flux quanta 
penetrating the 3-cycles $\Sigma \times \gamma'$, 
and are non-zero.
Thus, the Killing vectors of the vector fields 
(in D = 4 effective theory) $\int_{\gamma}C_{\gamma \mu}$ and 
$\int_{\gamma} B_{\gamma \mu}$ act non-trivially 
in the direction of the scalar $\int_{\Sigma \gamma\gamma'} C^{(4)}$. 
The introduction of fluxes turns on gauge coupling 
of the vector fields originating in the closed string sector.

The Ramond--Ramond scalars $\int_{\Sigma \gamma\gamma'} C^{(4)}$ are  
absorbed by the vector fields $\int_{\gamma}B_{\gamma \mu}$ 
through the Higgs mechanism in (\ref{eq:Stuckelberg}). 
The Fayet--Iliopoulos D-term parameters $e^3=\int_\Sigma \omega_{K3}$ 
are scalar ${\cal N}$ = 1 SUSY partners of the Ramond--Ramond scalars 
$\int_{\Sigma \gamma\gamma'} C^{(4)}$ (see Table \ref{tab:twoN=2}), 
and hence the D-term parameters are also stabilized by the fluxes 
as long as the ${\cal N}$ = 1 SUSY is preserved.
The Fayet--Iliopoulos F-term parameters 
$e^1+ie^2 = \int_\Sigma \Omega_{K3}$ are also stabilized when 
the ${\cal N}$ = 2 SUSY is preserved.
They are stabilized by the scalar potential (\ref{eq:result}), 
where they are contained in the momentum maps $P^x_\Lambda$.
The scalar partners of the vector fields, which are certain 
linear combinations of $X^\Lambda$'s ($\Lambda = 0,1,2,3$) 
in (\ref{eq:CV1}), 
also become massive when the ${\cal N}$ = 2 SUSY is preserved. 
Their mass term arises from the first term of 
(\ref{eq:fullpotential}), 
because $k^u_\Lambda$'s are non-zero for 
$u=\int_{\Sigma\gamma\gamma'}C^{(4)}$'s.

We introduce the fluxes so that the Killing vectors 
for $\Lambda = 2,3$ are turned on. 
This is because we do not want vacuum instability that arises 
due to the positive sign of $\eta^{\Lambda\Sigma}$ 
in (\ref{eq:result}). 
The Killing vectors we introduce later (and corresponding fluxes) 
preserve ${\cal N}$ = 2 SUSY. 
Thus, all the moduli mentioned above acquire masses. 

Other moduli, including the volume of $K3$ and $T^2$, 
are not stabilized in the toy model discussed 
in subsection \ref{ssec:quaternionic} and \ref{ssec:mixing}.
However, those moduli can be stabilized in the general framework 
of ${\cal N}$ = 1 supersymmetric vacua, and we just assume 
that they are stabilized at finite values and does not cause 
extra problems. Related discussion is found in subsection 
\ref{ssec:mixing} and \ref{ssec:CY}.

In subsection \ref{ssec:quaternionic}, 
we discuss in detail the potential stabilizing  
the Fayet--Iliopoulos parameters (blow-up modes) $e^m$'s ($m=1,2,3$).
The potential is roughly given by 
\begin{eqnarray}
 V & \sim & \frac{1}{2 \Im T} \left( 
   |P^x_{\Lambda=2}({\rm fcn.~of~}e^m\mbox{'s})|^2 
 + |P^x_{\Lambda=3}({\rm fcn.~of~}e^m\mbox{'s})|^2) \right) 
                        \nonumber \\
& & \qquad \qquad + \frac{1}{2 \Im S}|P^x_{\Lambda=4={\rm inflaton}}
        (e^x + {\rm fcn.~of~}(Q,\bar{Q})+\cdots)|^2 ,
\end{eqnarray}
where the first two terms arise from turning on non-trivial Killing
vectors for the bulk gauge fields, and the last term is for 
the gauge field on the fractional D3-brane. 
The first two terms fix the vacuum of $e^m$'s 
so that $P^x_2$ and $P^x_3$ vanish. 
The second line of (\ref{eq:result}), which is omitted here, 
also vanishes.
On the other hand, the effective Fayet--Iliopoulos parameters 
$P^x_4|_{Q,\bar{Q} = 0}$ do not vanish, 
because the function of $e^m$'s can be different for $P_{2,3}$ and 
for $P_4$, as we show explicitly in subsection 
\ref{ssec:quaternionic}. 
In particular, the positive
vacuum energy for the inflation is stabilized 
(when the volume of both $K3$ and $T^2$ are finite).
The purpose of subsection \ref{ssec:quaternionic} is 
to show explicitly that $P_{2,3}$ and $P_4$ 
can be different functions of the blow-up parameters. 

In subsection \ref{ssec:mixing}, we discuss the mixing of the inflaton 
with moduli $S$ and $U$ that is caused by the moduli stabilization. 
It turns out that there is no extra mass term generated by this mixing.
Although the inflaton mass does not vanish, 
we see that there is a flux configuration where the inflaton mass 
is sufficiently small.

\subsection{Quaternionic Geometry 
of the Hypermultiplets and \\ 
Stabilization of the Positive Vacuum Energy}\label{ssec:quaternionic}

The Fayet--Iliopoulos parameters are realized by VEV's of a 
hypermultiplet.
There are twenty hypermultiplets coming from the closed 
string sector, when the Type IIB theory is compactified 
on $K3 \times T^2/\Z_2$.
The eighty scalars consists of the moduli of $K3$ metric 
$e^{ma}$ ($m=1,2,3$, $a=1,...,19$) \cite{Aspinwall}, 
3+19 = 22 scalars $c^m$ ($m=1,2,3$) and $c^a$ ($a=1,...,19$) 
from the Ramond--Ramond 4-form, and $e^{-2\phi}$, which is 
the volume of $T^2$.
There are nineteen anti-self-dual 2-cycles in $K3$ manifold, 
and each of them has a triplet moduli $e^{ma}$ ($m=1,2,3$) 
describing the blow-up of the cycle.
The Fayet--Iliopoulos parameters we are interested in 
are $e^{ma}$ ($m=1,2,3$) for one of these cycles 
(one of $a \in \{1,...,19\}$).

In order to stabilize non-zero Fayet--Iliopoulos parameters, 
one has to know the quaternionic geometry for wider range 
of the moduli space, not just around the orbifold limit.
The global geometry of the quaternionic manifold is 
SO(4,20)/SO(4)$\times$SO(20) \cite{Seiberg}.
The global parametrization of this manifold, where coordinates 
are $(e^{ma},c^m,c^a,\phi)$, is explicitly described in \cite{AdAFL}.

Massless modes from the D3--D7 open string are also hypermultiplets,  
and thus, the total quaternionic
geometry is spanned by 80 coordinates of the bulk modes and 
extra coordinates of the open string modes.
The metric of the total quaternionic space is not known.  
However, the D3--D7 open string is given a large mass 
via (\ref{eq:hypermass}) and its VEV is zero during the inflation. 
Therefore, it is sufficient  to know the geometry of the 
submanifold where the VEV's of open string modes are zero, 
as long as we are concerned about the stabilization 
of the positive vacuum energy during the inflation.

We introduce the following Killing vectors:
\begin{eqnarray}
 k_{\Lambda=2} & = & g_1 \partial_{c^{m=1}} + g_2 \partial_{c^{a=1}}, 
   \qquad (g_1 < g_2),  \label{eq:rel-2-3A} \\
 k_{\Lambda=3} & = & g_1 \partial_{c^{m=2}} + g_2 \partial_{c^{a=2}}.
\label{eq:rel-2-3B}
\end{eqnarray}
The Killing vectors above are constant shifts in $c^m$ and $c^a$ 
directions, and it is easy to see that they are isometry;  
the metric of the quaternionic geometry is as follows:\change{
\begin{eqnarray}
ds^2 & = & d \phi ^2
+ \sum_m e^{2\phi} 
      ( \sqrt{1+e\cdot e^{t}}^{mn} dc^{n} + e^{ma}dc^{a})^2
+ \sum_a e^{2\phi} 
      ( dc^{m}e^{ma} +  dc^{b} \sqrt{1+e^{t} \cdot e}^{ba} )^2  
                          \nonumber \\
  & &  \qquad \qquad 
+ \sum_{a,m} ( \sqrt{1+e \cdot e^{t}}^{mn} de^{na} 
               - e^{mb} d\sqrt{1+e^{t} \cdot e}^{ba} )^2, \nonumber
\label{eq:metric}
\end{eqnarray}}{
\begin{eqnarray}
ds^2 & = & d \phi ^2
+ \sum_m e^{2\phi} 
      ( \sqrt{1+e\cdot e^{t}}^{mn} dc^{n} - e^{ma}dc^{a})^2
+ \sum_a e^{2\phi} 
      ( dc^{m}e^{ma} -  dc^{b} \sqrt{1+e^{t} \cdot e}^{ba} )^2  
                          \nonumber \\
  & &  \qquad \qquad 
+ \sum_{a,m} ( \sqrt{1+e \cdot e^{t}}^{mn} de^{na} 
               - e^{mb} d\sqrt{1+e^{t} \cdot e}^{ba} )^2, 
\label{eq:metric}
\end{eqnarray}}
which does not depend on $c^m$ and $c^a$.
This isometry is the remnant of the gauge symmetry 
adding an exact 4-form to $C^{(4)}$.
The ${\cal N}$ = 2 SUSY is preserved 
when the Killing vectors are chosen 
as in (\ref{eq:rel-2-3A}, \ref{eq:rel-2-3B}). 

The introduction of the Killing vectors (\ref{eq:rel-2-3A}) and 
(\ref{eq:rel-2-3B}) corresponds to introducing 3-form fluxes 
in the D = 10 picture. 
One can determine the fluxes in the D = 10 picture 
through (\ref{eq:Stuckelberg}), but 
we do not pursue this issue  further in this article.
The Killing vectors are sufficient information 
for the later purpose.

The Killing vectors are given, and now the momentum maps 
are obtained by \cite{Michelson,kp}
\begin{equation}
P^{x}_{\Lambda}=\omega^x_{u} k^u = 
\omega^{x}_{c^m}k_{\Lambda}^{c^m}+\omega_{c^a}^{x}k^{c^a}_{\Lambda}. 
\end{equation}
Here, $\omega^x$ is the su(2)$_R$ connection associated with 
the quaternionic manifold, which is given by 
\begin{equation}
 \omega^x = \omega ^{x}_{c^m}dc^{m}+\omega ^{x}_{c^a}dc^{a} + \cdots
  = e^{\phi} ( \sqrt{1+e \cdot e ^{t}}^{xm}dc^{m}
              - e^{xa}dc^{a}) + \cdots,
    \quad (x=1,2,3). 
\end{equation}
The ellipses stand for 1-form $d e^{ma}$ and $d\phi$.
Thus, the momentum maps are obtained:
\begin{eqnarray}
 P^x_{\Lambda=2} & = & e^\phi 
    \left(g_1 \sqrt{1+e \cdot e^t}^{x1} - g_2 e^{x1} 
    \right), \\
 P^x_{\Lambda=3} & = & e^\phi
     \left(g_1 \sqrt{1+e \cdot e^t}^{x2} - g_2 e^{x2} 
    \right).
\end{eqnarray}
All $e^{m1}$'s and $e^{m2}$'s are stabilized and their VEV's are 
determined 
by requiring the potential $(P^x_2)^2+(P^x_3)^2$ 
to be minimized. 
Their VEV's are 
\begin{eqnarray}
 && e^{11} = e^{22} = \frac{g_1}{\sqrt{g_2^2 - g_1^2}}, \\
 && \sqrt{1+(e^{11})^2} = \sqrt{1+(e^{22})^2} = 
      \frac{g_2}{\sqrt{g_2^2 - g_1^2}},  \\
 && e^{21}=e^{31}= e^{12}=e^{32}=0,  
\end{eqnarray}
and in particular, we see that the Fayet--Iliopoulos parameters 
can really be non-zero at the stabilized vacuum.

The Killing vector associated with $\Lambda=4$, i.e., the inflaton, 
is given by \change{\[
 k_{\Lambda=4} = g_3 \partial_{c^{a=1}} + 
  i \left(Q \partial_Q + \bar{Q}\partial_{\bar{Q}}\right) + {\rm h.c.},
\]}{
\begin{equation}
 k_{\Lambda=4} = g_3 \partial_{c^{a=1}} + 
  i \left(Q \partial_Q - \bar{Q}\partial_{\bar{Q}}\right) + {\rm h.c.},
\label{eq:KllgD3}
\end{equation} when the fractional D3 brane resides at the vanishing 2-cycle
corresponding to $c^{a=1}$,}
and \change{the momentum maps are roughly given by (\ref{eq:D3Dterm}) and 
(\ref{eq:D3Fterm}), where $e^m$'s in (\ref{eq:D3Dterm}) and 
(\ref{eq:D3Fterm}) are replaced by 
$e^\phi g_3 e^{m2}$.}
{the corresponding momentum map by $
P_{\Lambda=4}^x=-g_3 e^{\phi} e^{x1} + (\hbox{terms involving $Q, \tilde Q$}).
$} Thus, the positive vacuum energy 
$\vev{P^x_{\Lambda=4}P^x_{\Lambda=4}}$ is stable during the inflation 
(Here, $T$ and $e^{-\phi}$ are assumed to be stabilized \change{.}{by some other
mechanism}).
The vacuum energy is given by \change{
\[
 \rho_{\rm cos} = \frac{1}{\Im S} \left(e^\phi g_3 e^{22} \right)^2
                = \frac{e^{2\phi}}{\Im S}\frac{g_1^2}{g_2^2 - g_1^2}
                     g_3^2.
\]}{
\begin{equation}
 \rho_{\rm cos} = \frac{1}{\Im S} \left(e^\phi g_3 e^{11} \right)^2
                = \frac{e^{2\phi}}{\Im S}\frac{g_1^2}{g_2^2 - g_1^2}
                     g_3^2.
\label{eq:cos-cnst}
\end{equation}}

We have minimized $|P^x_{\Lambda=2}|^2$ and $|P^x_{\Lambda=3}|^2$ 
without considering the potential from $P_{\Lambda=4}$, and evaluated 
the potential from $P_{\Lambda=4}$ at the vacuum determined by 
$P_{\Lambda=2}$ and $P_{\Lambda=3}$.
\change{This treatment is justified}{The system dynamically minimizes
$|P^x_{\Lambda=2}|^2$ and $|P^x_{\Lambda=3}|^2$ }when 
the mass of the moduli $e^{ma}$ (denoted by $m_e$) is sufficiently 
larger than the Hubble parameter of the inflation 
$H \equiv \sqrt{\rho_{cos}}/(\sqrt{3}M_{\rm pl}) 
\simeq \sqrt{\rho_{cos}}$, i.e.,\change{
\[
\frac{H^2}{m_e^2} \sim \frac{g_3^2 \vev{e}^2/\Im S}
                            {((g_2^2-g_1^2)/g_2)^2 / \Im T / \vev{e}^2}
   \sim \frac{g_3^2 {\rm vol}(K3)}
             {(g_2^2-g_1^2)^4/(g_1^4 g_2^2)} \lsim 1.
\]}{\begin{equation}
\frac{H^2}{m_e^2} \sim \frac{e^{2\phi}g_3^2 \vev{e}^2/\Im S}
                            {e^{2\phi}((g_2^2-g_1^2)/g_2)^2 / \Im T}
   \sim \frac{g_3^2 {\rm vol}(K3)}
             {(g_2^2-g_1^2)^3/(g_1^2 g_2^2)} \lsim 1,
\label{eq:ratioHMe}
\end{equation} where we have assumed $\vev{e}\ll 1$.}
\change{This is also a necessary condition for a inflation model. 
Otherwise the value of the Fayet--Iliopoulos parameters 
would change considerably along the evolution of the inflation. }{
Although the above picture---$P_{\Lambda=2,3}=0$ and $\rho_{\rm cos}$ 
is given by Eq.~(51) during inflation---is slightly modified in section
4.2, it is shown that the inflation dynamics is not affected 
essentially.
The value of moduli determined in Eqs.~(47--49) and the vacuum energy 
Eq.~(51) are used as the first order approximation during inflation. 
}


\subsection{Inflaton--Moduli Mixing  
and Slow-roll Conditions}\label{ssec:mixing}

We have assumed so far that the first term 
in (\ref{eq:fullpotential}) does not play an important role.
This term, however, contains a potential corresponding to 
the quadratic term in (\ref{eq:crude}), and hence can be 
harmful to the evolution of the inflaton.
Therefore, let us now turn our attention to this term 
and determine what circumstances it is not harmful.

The vacuum of the hypermultiplets is determined from 
the potential (\ref{eq:result}) in the previous subsection. 
Now it turns out that $\vev{h_{uv}k^u_\Lambda k^v_\Sigma}$ does not 
vanish. Thus, this term generates mass terms to the scalar 
particles in the vector multiplets. The mass term is given by
\change{
\begin{eqnarray}
& & \frac{e^{2\phi}}{\Im S \Im T \Im U} 
\left(
  \left( (g_1 X^2)^\dagger, (g_2 X^2)^\dagger \right) 
  \left(\begin{array}{cc}
         c^2 + s^2 & 2 s c \\
         2 s c & c^2 + s^2 \\
        \end{array}\right)
  \left(\begin{array}{c}
          g_1 X^2 \\ g_2 X^2
        \end{array}\right) 
\right. \nonumber \\
 & & \left.+  \left( (g_1 X^3)^\dagger, (g_2 X^3 + g_3 X^4)^\dagger 
              \right) 
 \left(
 \begin{array}{cc}
   c^2 + s^2 & 2 s c \\
   2 s c & c^2 + s^2 \\
 \end{array}
 \right)
  \left(
  \begin{array}{c}
      g_1 X^3 \\ g_2 X^3 + g_3 X^4
  \end{array}
  \right)
\right),\nonumber
\end{eqnarray}}{\begin{eqnarray}
& & \frac{e^{2\phi}}{\Im S \Im T \Im U} 
\left(
  \left( (g_1 X^3)^\dagger, (g_2 X^3)^\dagger \right) 
  \left(\begin{array}{cc}
         c^2 + s^2 & -2 s c \\
         -2 s c & c^2 + s^2 \\
        \end{array}\right)
  \left(\begin{array}{c}
          g_1 X^3 \\ g_2 X^3
        \end{array}\right) 
\right. \nonumber \\
 & & \left.+  \left( (g_1 X^2)^\dagger, (g_2 X^2 + g_3 X^4)^\dagger 
              \right) 
 \left(
 \begin{array}{cc}
   c^2 + s^2 & -2 s c \\
   -2 s c & c^2 + s^2 \\
 \end{array}
 \right)
  \left(
  \begin{array}{c}
      g_1 X^2 \\ g_2 X^2 + g_3 X^4
  \end{array}
  \right)
\right),
\label{eq:quadratic}
\end{eqnarray}}
where $X^{\Lambda=2,3,4}$ are those in (\ref{eq:CV1}) and 
abbreviated notations $c^2 \equiv g_2^2 / (g_2^2 - g_1^2)$ 
and $s^2 \equiv g_1^2 / (g_2^2 - g_1^2)$ are introduced.
Here, the metric (\ref{eq:metric}), the Killing vectors
(\ref{eq:rel-2-3A}, \ref{eq:rel-2-3B}, \ref{eq:KllgD3}) and 
the K\"{a}hler potential (\ref{eq:Kahler2}) are used along with 
(\ref{eq:LX}).
The first line of the above potential leads
 \change{$X^{\Lambda=2}$}{$X^{\Lambda=3}$} to zero.
The mass matrix in the second line is diagonalized; \change{
\begin{eqnarray}
{\rm eigenvalue:} \; (c+s)^2 = \frac{g_2+g_1}{g_2-g_1}, & &
{\rm eigenstate:} \; \frac{1}{\sqrt{2}} ((g_2+g_1)X^3 + g_3 X^4), \nonumber\\
{\rm eigenvalue:} \; (c-s)^2 = \frac{g_2-g_1}{g_2+g_1}, & &
{\rm eigenstate:} \; \frac{1}{\sqrt{2}} ((g_2-g_1)X^3 + g_3 X^4).\nonumber
\end{eqnarray}
The former eigenstate is dominantly $X^{\Lambda=3}$
since $g_3 \ll g_1,g_2$, and the other 
has (mass)$^2$ eigenvalue suppressed by $(g_2-g_1)/(g_2+g_1)$.
}{
\begin{eqnarray}
{\rm eigenvalue:} \; (c-s)^2 = \frac{g_2-g_1}{g_2+g_1}, & &
{\rm eigenstate:} \; \frac{1}{\sqrt{2}} ((g_2+g_1)X^2 + g_3 X^4), \\
{\rm eigenvalue:} \; (c+s)^2 = \frac{g_2+g_1}{g_2-g_1}, & &
{\rm eigenstate:} \; \frac{1}{\sqrt{2}} ((g_2-g_1)X^2 + g_3 X^4).
\end{eqnarray}
}

\change{The (mass)$^2$ of $X^2$ and $(X^3 + g_3/(g_2+g_1) X^4)$ are not smaller 
than the squared Hubble parameter because 
\[
 m^2 \gsim \frac{e^{2\phi}}{\Im S}g_1^2\frac{g_2+g_1}{g_2-g_1}
    \gsim \frac{e^{2\phi}}{\Im S}g_3^2\frac{g_1^2}{g_2^2-g_1^2}=H^2.
\]}{
The (mass)$^2$ of $X^3$ and $(X^2 + g_3/(g_2+g_1) X^4)$ are not smaller 
than the squared Hubble parameter because 
\begin{equation}
 m^2 \gsim \frac{e^{2\phi}}{\Im T}(g_2^2-g_1^2)\sim
 \frac{g_2^2}{g_2^2-g_1^2}m_e^2     \gsim H^2
\end{equation}}
Thus, the moduli $S$ and $U$ are determined by \change{
\[
X^2 = 0, \quad {\rm and}\quad X^3 = - \frac{g_3}{g_2+g_1}X^4
\]}{
\begin{equation}
X^3 = 0, \quad {\rm and}\quad X^2 = - \frac{g_3}{g_2+g_1}X^4
\label{eq:mixing}
\end{equation}}
as functions of the inflaton  $X^{\Lambda=4}=Z$.
In particular,\change{\[
 \Im S = 1 - \left(\frac{g_3}{2(g_1+g_2)}Z\right)^2 + \cdots,
\]}{ 
\begin{equation}
 \Im S = 1 + \left(\frac{g_3}{2(g_1+g_2)}|Z|\right)^2 + \cdots,
\label{eq:stable-gs}
\end{equation}}
where $Z$ is assumed to be \change{real}{purely imaginary} for simplicity. 

The moduli $S$ and $U$ are integrated out, i.e., the relations 
(\ref{eq:mixing}) are substituted into the potential 
(\ref{eq:quadratic})+(\ref{eq:cos-cnst}).
The net effect of integrating out heavy moduli is to replace 
$\Im S$ with (\ref{eq:stable-gs}) in (\ref{eq:cos-cnst}) and 
the original inflaton $X^4=Z$ with a linear combination of 
\change{$X^3$}{$X^2$} and $X^4$ in (\ref{eq:quadratic}). 
\change{The inflaton $Z$ is canonically normalized, and 
now }{After canonically normalizing the inflaton $Z$,}
we finally obtain the total effective action 
relevant to the inflation\change{:
\[ {\cal L} \simeq M_{\rm pl}^2 \left( | \partial \tilde Z|^2 
    - \left(1 + \frac{g_3^2 {\rm vol}(K3)}{4 (g_1+g_2)^2}\tilde Z^2 
      \right) e^{2\phi} 
      \left(
        \frac{1}{2}\frac{g_2 - g_1}{g_2 + g_1}
        \left| \frac{2g_1}{g_2+g_1}g_3 \tilde Z  \right|^2 
      + \frac{g_1^2}{g_2^2 - g_1^2}g_3^2 
      \right)
                               \right),
\]}{
\begin{equation}
 {\cal L} \simeq M_{\rm pl}^2 \left( | \partial \tilde Z|^2 
    - \left(1 - \frac{g_3^2 {\rm vol}(K3)}{4 (g_1+g_2)^2}|\tilde Z|^2 
      \right) e^{2\phi} 
      \left(
        \frac{1}{2}\frac{g_2 + g_1}{g_2 - g_1}
        \left| \frac{2g_1}{g_2+g_1}g_3 \tilde Z  \right|^2 
      + \frac{g_1^2}{g_2^2 - g_1^2}g_3^2 
      \right)
                               \right),
\label{eq:final}
\end{equation}}
where $\tilde Z$ is the canonically normalized inflaton.
Thus, the slow-roll condition (\ref{eq:sr}) implies that \change{
\[
 \eta \simeq \frac{1}{2}\left(\frac{g_2-g_1}{(g_2+g_1)/2}\right)^2 
     + \frac{g_3^2 {\rm vol}(K3)}{4(g_1+g_2)^2} \ll 1. 
\]
The first term is sufficiently small when 
the two flux quanta $g_1$ and $g_2$ are degenerate by 10 \%.
Under this condition, the blow-up parameters 
of the $K3$ manifold are  
\[
 e^{11}=e^{22}=\sqrt{\frac{g}{g_2-g_1}} \gsim (2-3),
\]
where $g_1 \sim g_2 \sim g$.
The second term, which comes from the inflaton dependence 
of the string coupling, is sufficiently small when 
the above condition and (\ref{eq:ratioHMe}) are satisfied. 
}{
\begin{equation}
 \eta \simeq 2
     - \frac{g_3^2 {\rm vol}(K3)}{4(g_1+g_2)^2} \ll 1. 
\end{equation}
We find that the purely imaginary direction of $Z$ is
sufficiently flat when $g_3^2\vol{(K3)}\simeq 8(g_1+g_2)^2$.
This requirement for the slow-roll potential
 is compatible with Eq. (52) when\begin{equation}
(e^{11})^2=(e^{22})^2=\frac{g_1^2}{g_2^2-g_1^2} \lsim \frac18
\end{equation}
}

\change{It is also easy to see that the other slow-roll parameter $\epsilon$ 
is also sufficiently small under the above condition. 
The second line in (\ref{eq:result}), which has 
linear dependence on the inflaton, contributes to $\epsilon$ by 
\[
 \epsilon \sim \frac{H^2}{m_e^2} \frac{\Im S}{\vev{e}^4} \lsim \eta.
\]}
{It is also easy to see that the other slow-roll parameter $\epsilon$ 
is also small. It is because the term linear in the inflaton $y$ in Eq. (28)
vanishes since it is multiplied by $P_{\Lambda=3}=0$.
It should be noted that $e^{22}$ appears only in $|P_{\Lambda=3}|^2$ but
not in $|P_{\Lambda=2}|^2$ or $|P_{\Lambda=4}|^2$, and hence 
$P_{\Lambda=3}$ is zero even during inflation.
Although the value of $e^{11}$ during inflation is shifted from 
that given in Eq.~(47), the only effects of this shift are to reduce $\rho_{\rm cos}$, 
$H^2$ and to increase $m_e$. 
Thus, all the discussion so far is still valid.
The modulus $e^{11}$
slides to the vacuum value when the hybrid inflation ends and $P_{\Lambda=4}$
vanishes. Since it has mixing with the inflaton, it can decay to light particles 
through the mixing, and the evolution of $e^{11}$ does not lead to a serious
cosmological moduli problem.}

It has been assumed so far that the volume of the torus $e^{-2\phi}$ 
does not have $Z$-dependence. If it were stabilized as functions 
of the inflaton $Z$, the inflaton potential in (\ref{eq:final}) 
would be no longer flat.
Therefore, the conditions for the slow-roll inflation are  
i) the $T^2$ volume is stabilized independently from $Z$,  
\change{ii) the flux quanta $g_1$ and $g_2$ are degenerate by 10\%, and 
iii) the flux  quanta $g_1 \sim g_2$ are sufficiently large so that 
the moduli mass is larger than the Hubble parameter.}{
ii) $(g_1/g_2)^2\lsim 1/8$
and  iii) $g^2_3\vol(K3)\simeq 8(g_1+g_2)^2$ at a few
percent level.
}

The volume of 
the torus $e^{-2\phi}$ is irrelevant to the slow-roll condition.
Thus, it can be arbitrary (from the view point of phenomenology), 
and in particular, can be moderately large 
so that the exponential terms in (\ref{eq:crude}) are 
sufficiently suppressed.

\change{The two moduli $S$ and $U$ change through (\ref{eq:mixing}) 
as the value of the inflaton $Z$ changes. However, the slow-roll 
condition implies that the resulting changes of $S$ and $U$ 
are not significant.}{}

Finally, one remark is in order here.
The coordinate of the D3-brane $Z=X^4$ explicitly appears 
in the scalar potential, and it looks as if 
the origin of the torus has a physical meaning.
This is actually an artifact of our treatment, where we focused 
only on one fractional D3-brane.
When all the D-branes relevant to the twisted RR-charge 
cancellation are introduced, we expect 
that the potential will be a function only of the distance between 
those D-branes. We consider that ``$Z$'' we used in this article 
is an approximation, in some sense, to the distance between 
the fractional D3-brane and one of those D7-branes. 

\subsection{Moduli Stabilization 
in Generic Calabi--Yau Manifold}\label{ssec:CY}

\begin{table}[t]
\begin{center}
 \begin{tabular}{c|c|c}
  & ${\cal N}$ = 2 hypermultiplet & ${\cal N}$ = 2 vector multiplet 
     \\
  &  of $K3 \times T^2/\Z_2$ & of $K3 \times T^2/\Z_2$ \\
\hline
${\cal N}$ = 2 hypermultiplet of $CY_3$ & 
  $\int_{\Sigma}C^{(4)}_{\Sigma \mu\nu}$, 
  $e^3 = \int_\Sigma \omega_{K3}$ & 
  $\int_{\Sigma} B,\int_\Sigma C^{(2)}$ \\
\hline
${\cal N}$ = 2 vector multiplet of $CY_3$ & 
  $e^1+i e^2 = \int_{\Sigma \gamma } \Omega$ &
  $\int_{\Sigma \gamma} C^{(4)}$
 \end{tabular}
\caption{\label{tab:twoN=2}(Some of) Moduli particles are classified 
in terms of two different ${\cal N}$ = 2 SUSY. Particles in the 
right column are odd under the orientifold projection $\Z_2$, 
and are projected out. $\Sigma$ stands for a 2-cycle in $K3$ and 
$\gamma$ for an 1-cycle of $T^2$. $\Omega$ is the 
global holomorphic 3-form of $CY_3$.}
 \end{center}
\end{table}

Some of the results obtained in subsection \ref{ssec:quaternionic} 
and \ref{ssec:mixing} are specific to choice of  
$K3 \times T^2$ as the Calabi--Yau 3-fold.  
Thus, we go back to the most generic setup described at the beginning 
of section \ref{sec:setup}, where the Calabi--Yau 3-fold is required 
only to have local geometry ALE $\times$ {\bf C}, and discuss 
issues relevant to the moduli stabilization again.

Let us start with the Type IIB theory compactified on a 
Calabi--Yau 3-fold without space-filling D-branes. 
Moduli particles are classified into ${\cal N}$ = 2 SUSY 
multiplets\footnote{
Note that the eight SUSY charges of this ${\cal N}$ = 2 SUSY are 
not the same subset of the 32 SUSY charges of the Type IIB theory 
as those of the ${\cal N}$ = 2 SUSY in section \ref{sec:4DSUGRA}. 
Only 4 SUSY charges (${\cal N}$ = 1 SUSY) belong to the both. 
See Table \ref{tab:twoN=2}.}.
There are $h^{2,1}$ vector multiplets 
$(\int_A \Omega, \int_A C^{(4)}_{A\mu})$ and 
$h^{1,1}$ hypermultiplets 
$((\int_\Sigma C^{(4)}_{\Sigma \mu\nu},\int_\Sigma \omega),
  (\int_\Sigma B,\int_\Sigma C^{(2)}))$, where $A$ and $\Sigma$ 
denote 3-cycles and 2-cycles of the Calabi--Yau 3-fold, respectively.
There is another hypermultiplet 
$((S \equiv C^{(0)}+ie^{-\phi}),(B_{4D},C^{(2)}_{4D}))$.
When 3-form fluxes and O3-planes are introduced, only ${\cal N}$ = 1 
SUSY can be  preserved, and ${\cal N}$ = 1 multiplets 
$\int_A C^{(4)}_{A\mu}$, $(\int_\Sigma B,\int_\Sigma C^{(2)})$ 
and $(B_{4D},C^{(2)}_{4D})$ are projected out. 
The ${\cal N}$ = 1 chiral multiplets $\int_A \Omega$ are stabilized 
by effective superpotential induced by fluxes \cite{GVW} 
\begin{eqnarray}
 W & = & \int_{CY_3} \Omega \wedge G = 
   \int_A \Omega \vev{\int_B G} - \int_B \Omega \vev{\int_A G}, \\
 G & \equiv & d C^{(2)} - S d B,
\label{eq:Gukov}
\end{eqnarray}
where $\int_B \Omega$'s are written as functions of 
$\int_A \Omega$'s.
Thus, in particular, the Fayet--Iliopoulos F-term 
$e^1+ie^2 = \int_\Sigma \Omega_{K3} 
= \int_{\Sigma \gamma}\Omega_{CY_3}$ 
and the chiral multiplet $S$ 
are stabilized by this superpotential. 
The stable minimum of $\int_A \Omega_{CY_3}$ depends on 
the fluxes introduced, and can be non-zero.
On the other hand, the Fayet--Iliopoulos D-term parameter 
$e^3 = \int_\Sigma \omega_{K3} = \int_\Sigma \omega_{CY_3}$ 
is not stabilized through this superpotential (\ref{eq:Gukov}). 
But, non-perturbative effects of gauge theories might help stabilizing 
these moduli. 

It is surely possible that all the moduli are stabilized and 
that the effective Fayet--Iliopoulos parameters are non-zero. 
However, this is not enough for the model of inflation.
Let us suppose that the moduli stabilization in (\ref{eq:Gukov})
is effectively described by the following superpotential
\begin{equation}
W_{\rm moduli} = 
M_0 + M_2 (\Xi - \zeta^2)^2 + {\cal O}((\Xi - \zeta^2)^3),
\label{eq:moduli-eff}
\end{equation} 
where $\Xi$ denotes a modulus chiral multiplet whose VEV provides 
the Fayet--Iliopoulos parameter, and $M_0$, $M_2$ and $\zeta$ 
are numerical parameters.
Then, the total system is governed by
\begin{equation}
W = \sqrt{2}g X(Q\bar{Q} - \Xi) + W_{\rm moduli},
\end{equation}
and the effective superpotential obtained after the modulus $\Xi$ 
is integrated out contains a mass term of the inflaton $X$. 
Thus, the inflaton potential is no longer flat.

This is not the case when the effective model of 
the moduli stabilization (\ref{eq:moduli-eff}) is replaced by 
\begin{equation}
 W_{\rm moduli} = X'  \; \; {\rm fcn.}( \Xi ),
\label{eq:Rinv-moduli}
\end{equation}
where $X'$ is another modulus. 
One linear combination of $X$ and $X'$ 
is integrated out, while the other combination remains light, 
and plays the role of the inflaton.
The toy model of the moduli stabilization given 
in subsection \ref{ssec:quaternionic} and \ref{ssec:mixing} 
is partly described by this superpotential; 
$X^{\Lambda=3}$ plays the role of $X'$.

One of remarkable features of the hybrid inflation model 
\cite{F-SUSY-a,F-SUSY-b,F-SUSY-c,D-SUSY} is that 
there is a (discrete) $R$ symmetry, 
under which $X$ carries $R$ charge 2 \cite{F-SUSY-b,WY}.
Thus, if there is a moduli stabilization that preserves 
such a (discrete) $R$ symmetry, as in the superpotential 
(\ref{eq:Rinv-moduli}), 
the effective superpotential of the inflaton is still 
constrained by the $R$ symmetry even after the moduli 
are integrated out, and the inflaton potential remains flat. 
Therefore, the string realization of the $R$-invariant 
moduli stabilization deserves further investigation.

\section*{Acknowledgements} 
The authors thank S.~Yamaguchi and T.~Yanagida for discussion.
This work is supported in part (TW) by 
Japan Society for the Promotion of Science, 
Miller Institute for Basic Research of Science, and 
the Director, Office of Science, Office of High Energy and 
Nuclear Physics, of the U.S.Department of Energy under Contract 
DE-AC03-76SF00098.

\end{document}